\documentclass[aip,reprint,nofootinbib,amsmath,amssymb,longbibliography,floatfix,10pt,tightenlines,superscriptaddress]{revtex4-2}

\usepackage[english]{babel}
\usepackage{bm,graphicx,braket,color,booktabs,fixmath,dsfont}
\usepackage[hidelinks,colorlinks=true,citecolor=black,linkcolor=black,urlcolor=black]{hyperref}
\usepackage{algorithm}
\usepackage[noend]{algpseudocode}
\usepackage{hyperref}
\include{lang_fix}

\begin{document}

\title{Massively parallel pixel-by-pixel nanophotonic optimization using a Green's function formalism}

% \author{Jiahui Wang$^*$}
% % \thanks{These authors contributed equally to this work.}
% \affiliation{X, 100 Mayfield Ave, Mountain View, CA 94043, USA}
% \affiliation{Department of Applied Physics, Stanford University, 348 Via Pueblo Mall, Stanford, CA 94305, USA}
% % \email{jiauiw@stanford.edu}
% \author{Alfred K. C. Cheung$^*$}
% % \thanks{These authors contributed equally to this work.}
% \email{alfredkcc@x.team}
% \affiliation{X, 100 Mayfield Ave, Mountain View, CA 94043, USA}
% \author{Aleksandra Spyra}
% \affiliation{X, 100 Mayfield Ave, Mountain View, CA 94043, USA}
% \author{Ian A. D. Williamson}
% \affiliation{X, 100 Mayfield Ave, Mountain View, CA 94043, USA}
% \author{Jian Guan}
% \affiliation{X, 100 Mayfield Ave, Mountain View, CA 94043, USA}
% \author{Martin~F.~Schubert}
% \affiliation{X, 100 Mayfield Ave, Mountain View, CA 94043, USA}
% \def\thefootnote{*}\footnotetext{These authors contributed equally to this work}\def\thefootnote{\arabic{footnote}}

\author{Jiahui Wang,$^{1,2,*}$ Alfred K. C. Cheung,$^{1,*,\ddagger}$ Aleksandra Spyra,$^{1}$ Ian A. D. Williamson,$^{1}$ Jian Guan,$^{1}$ and Martin~F.~Schubert$^{1}$}
\noaffiliation{}
\affiliation{X, 100 Mayfield Ave, Mountain View, CA 94043, USA}
\affiliation{Department of Applied Physics, Stanford University, 348 Via Pueblo Mall, Stanford, CA 94305, USA}
\affiliation{X, 100 Mayfield Ave, Mountain View, CA 94043, USA}
\def\thefootnote{*}\footnotetext{These authors contributed equally to this work.}\def\thefootnote{\arabic{footnote}}
\def\thefootnote{$\ddagger$}\footnotetext{Corresponding author: alfredkcc@x.team}\def\thefootnote{\arabic{footnote}}

\newcommand{\vect}[1][\mathbf]{#1}
\newcommand{\dprime}{{\prime\prime}}

% \date{\today}

\begin{abstract}
We introduce an efficient parallelization scheme to implement pixel-by-pixel nanophotonic optimization using a Green's function based formalism. The crucial insight in our proposal is the reframing of the optimization algorithm as a large-scale data processing pipeline, which allows for the efficient distribution of computational tasks across thousands of workers. We demonstrate the utility of our implementation by exercising it to optimize a high numerical aperture focusing metalens at problem sizes that would otherwise be far out of reach for the Green's function based method. Finally, we highlight the connection to powerful ideas from reinforcement learning as a natural corollary of reinterpreting the nanophotonic inverse design problem as a graph traversal enabled by the pixel-by-pixel optimization paradigm.
\end{abstract}

\maketitle

\section{Introduction}

The promise of photonic inverse design~\cite{Molesky1:18} is to enable the optimization of non-intuitive photonic structures that achieve superior device performance (e.g. lower losses, larger bandwidths) within substantially more compact device footprints. This requires the effective exploration of complex and high dimensional design spaces to ultimately arrive at designs which are both performant and fabricable by modern foundry manufacturing processes. The fabricability requirements are crucial and broadly entail the enforcement of: (a) a binary condition on the value of the permittivity at each design pixel (assuming, as in typical examples, two possible materials), and (b) compliance with minimum feature size constraints for printed features as required by manufacturing processes. As such, optimization schemes that are able to maintain one or both fabricability requirements throughout the entire course of an optimization are especially attractive.

An always-feasible design is notably \textit{not} a property of the most widely used class of optimization methods for photonic inverse design, which rely on continuous optimization algorithms and the adjoint variable method (AVM)~\cite{Wang1:18,Veronis1:04,Molesky1:18,Hammond1:21,Sell1:17}. In the AVM, changes to the design are made based on the gradient of the device figure of merit (FOM) with respect to the permittivity at each design pixel. The gradient provides the response of the FOM with respect to infinitesimally small, continuous changes in the permittivity at each pixel. The gradient at a pixel will not, in general, reflect the actual change in the FOM that results from discrete, arbitrarily large changes in the permittivity. Therefore, while formalisms have been developed that utilize the gradient in ways that try to drive the design towards fabricability~\cite{Piggott1:15,Vercruysse1:19,Hammond1:21,Sell1:17,Piggott2:17}, there are no strict guarantees, and there is typically a strong trade off between driving towards fabricability and device performance.

An alternative optimization paradigm is to instead frame the optimization problem as a pixel-by-pixel graph traversal scheme, where every node of a graph is a possible design. Specifically, the root node design could be a featureless design, and the children of any design are all possible modifications to the design. Conceptually, as an implementation of this strategy, one might at each optimization step simulate a set of designs that result from discrete variations in the permittivity at one or multiple pixels, calculate the resulting changes in FOM, and greedily update the design to the variation achieving the largest positive change in FOM. The pixel-by-pixel paradigm possesses the advantage of always satisfying the binary condition. Furthermore, by imposing additional rules on how the actions are selected, minimum feature size design rules can also be easily enforced as a byproduct of the graph traversal~\cite{Schubert1:22}. Finally, the reinterpretation of the photonic inverse design problem as a graph traversal naturally lends itself to application of ideas from reinforcement learning (RL)---a strategy which has proven to be quite fruitful across a broad spectrum of domains in recent years~\cite{Silver1:18,Mirhoseini1:21}. We seek to showcase this idea with a demonstration later in this work.

A brute force implementation of the pixel-by-pixel optimization paradigm based on numerous (full-wave) simulations of the design variations at each optimization step is extremely computationally intensive and is limited to the optimization of small systems~\cite{Shen1:15}. In Ref.~\onlinecite{Boutami1:19}, Boutami and Fan introduce a pixel-by-pixel optimization formalism based on the Green's function technique~\cite{Martin1:94,Martin2:98} for solving the electromagnetic scattering problem that circumvents the need for any full-wave simulations. The authors demonstrate that if the Green's function is known for a given structure, the change in FOM for discrete variations of the structure can be efficiently evaluated. Subsequently, after a design variation is selected based on the change in FOM information, the Green's function can also be updated efficiently. Thus, as long as the Green's function for the initial structure is known, the pixel-by-pixel optimization scheme can proceed without the need for any full-wave simulations.

The Green's function formalism is difficult to scale to even moderately sized systems due to the formidable memory cost (scaling as $\mathcal{O}(N^2)$ where $N$ is the total number of pixels in the design region) of storing the full two-point Green's function throughout the optimization. The authors of Ref.~\onlinecite{Boutami1:19} recognized this and concurrently introduced a modified formalism in Ref.~\onlinecite{Boutami2:19} where, instead of having to store the full Green's function for all pairwise combinations of positions, only the \textit{position diagonal elements} are required in exchange for a small number of full-wave simulations needed at each optimization step. This drastically reduces the memory cost to $\mathcal{O}(N)$. This alternative implementation was exercised to optimize for 3D silicon-on-insulator waveguide bends which were fabricated and measured in Ref.~\onlinecite{Boutami3:20}.

While the memory-reduced alternative implementation of Ref.~\onlinecite{Boutami2:19} is undoubtedly promising, it comes with the cost of additional computational complexity associated with the reintroduction of full-wave simulations at each step. Indeed, there is a unique appeal to the original implementation of Ref.~\onlinecite{Boutami1:19} in forgoing any full-wave simulations. This sets the formalism apart from practically all other topology optimization schemes.

In this work, we demonstrate that it is possible to scale to large design problems ($\sim 10^6$ design pixels) using the formalism introduced in Ref.~\onlinecite{Boutami1:19} in spite of the $\mathcal{O}(N^2)$ memory scaling for storing the full Green's function. The crucial insight is the observation that the structure of the optimization scheme fits naturally within large scale data processing models~\cite{Dean1:04,Akidau1:15} wherein the storage, change in FOM calculations, and Green's function updates are all trivial to chunk and parallelize across thousands of workers or more on computer clusters. We will show that our massively parallel implementation of the Green's function based pixel-by-pixel optimization scheme is a viable option for photonic inverse design. We further provide reasons for why it is particularly well suited for application to the inverse design of metalenses.

This paper is organized as follows. In Sec.~\ref{sec:formalism}, we review the formalism behind the Green's function pixel-by-pixel optimization method. In Sec.~\ref{sec:parallelization}, we then demonstrate how the optimization scheme can be massively parallelized through an implementation that uses common data processing frameworks such as the open source Apache Beam~\cite{Akidau1:15,Apache:21} which conforms to the Dataflow~\cite{43864} model. In Sec.~\ref{sec:metalens}, we turn to an application of our implementation to the design of a high numerical aperture (NA) metalens. Section~\ref{sec:summary} provides a summary and further discussion of the results, with a particular emphasis on the connections of pixel-by-pixel optimization with fundamental concepts in RL.

\section{Formalism} \label{sec:formalism}

Consider a three-dimensional photonic inverse design problem in which the design degrees of freedom are the discrete set of permittivity values (corresponding to a set of possible materials) at $N$ pixel locations within a design region $\mathcal{D}$, where each pixel has volume $\Delta V$. For simplicity, let us restrict our attention to the case of two materials. We refer to the first material as the \textit{patterning material} with relative permittivity $\varepsilon$ and the second material as the \textit{background material} with relative permittivity $\varepsilon_0$, and define $\Delta\varepsilon\equiv\varepsilon - \varepsilon_0$. Let $\lambda$ be the operating wavelength in the background material and $k_0 = 2\pi / \lambda$. Finally, let the FOM be a scalar function $F$ of the electric field at $N_m$ monitor positions $\{\vect{r}_{m,i}: i=1,...,N_m\}$:
\begin{align}
\textrm{FOM} = F(\vect{E}(\vect{r}_{m,1}),...,\vect{E}(\vect{r}_{m,N_m})).
\label{eq:form1}
\end{align}
The goal of the optimization is to maximize the value of the FOM by sequentially proposing discrete modifications to the design and selecting the modifications based on exact $\Delta \textrm{FOM}$ values. The main idea behind the Green's function formalism is that, if the Green's function $\vect{G}(\vect{r},\vect{r}^\prime)$ is known for all pairs of positions $\vect{r},\vect{r}^\prime \in \mathcal{D} \cup \{\vect{r}_{m,i}\}$ for a structure, then $\Delta \textrm{FOM}$ values for single or multiple pixel modifications to the structure can be calculated efficiently without the need for any full-wave simulations.

The starting point is the Green's function technique for solving the electromagnetic scattering problem~\cite{Martin1:94,Martin2:98}. Consider a reference structure (i.e. a distribution of background and patterning material) for which the Green's function $\vect{G}_0(\vect{r},\vect{r}^\prime)$ is known. The Green's function $\vect{G}(\vect{r},\vect{r}^\prime)$ corresponding to a modification of the reference structure where background material in a region $V\subset\mathcal{D}$ is flipped to patterning material can be self-consistently related to $\vect{G}_0(\vect{r},\vect{r}^\prime)$ via Dyson's equation~\cite{Boutami1:19,Martin1:94}:
\begin{align}
\vect{G}(\vect{r},\vect{r}^\prime) = \vect{G}_0(\vect{r},\vect{r}^\prime) + \int_V d\vect{r}^\dprime \vect{G}_0(\vect{r},\vect{r}^\dprime)\cdot k_0^2 \Delta\varepsilon \vect{G}(\vect{r}^\dprime,\vect{r}^\prime).
\label{eq:form2}
\end{align}
The electric fields for the reference structure, $\vect{E}_0(\vect{r})$, and for the modified structure, $\vect{E}(\vect{r})$, obey the Lippman-Schwinger equation~\cite{Boutami1:19,Martin1:94}:
\begin{align}
\vect{E}(\vect{r}) = \vect{E}_0(\vect{r}) + \int_V d\vect{r}^\dprime \vect{G}_0(\vect{r},\vect{r}^\dprime)\cdot k_0^2 \Delta\varepsilon \vect{E}(\vect{r}^\dprime).
\label{eq:form3}
\end{align}

\subsection{Calculating changes in the FOM} \label{subsec:deltafom}

We now reinterpret Eqs.~(\ref{eq:form2}) and~(\ref{eq:form3}) within the context of pixel-by-pixel optimization by regarding $\vect{E}_0(\vect{r})$ and $\vect{G}_0(\vect{r},\vect{r}^\prime)$ as the \textit{known} field and Green's function for the structure at the start of a given optimization step. We propose a modified structure where pixels at positions $\vect{r}^\dprime$ within a region $V$ are flipped. Discretizing Eq.~(\ref{eq:form3}) and hence converting the integral into a summation, the field that results from this structural modification is:
\begin{align}
\vect{E}(\vect{r}) = \vect{E}_0(\vect{r}) + b\sum_{\vect{r}^\dprime_i \in V}\vect{G}_0(\vect{r},\vect{r}^\dprime_i)\cdot \vect{E}(\vect{r}^\dprime_i),
\label{eq:form4}
\end{align}
where, for convenience, we have defined the constant $b\equiv k_0^2 \Delta\varepsilon\Delta V$. To solve Eq.~(\ref{eq:form4}) for the field at general positions $\vect{r}$, we must first solve for the fields at the positions of the flipped pixels in $V$. Evaluating the field at the position $\vect{r}^\dprime_j\in V$, we obtain:
\begin{align}
\vect{E}(\vect{r}^\dprime_j) = \vect{E}_0(\vect{r}^\dprime_j) + b\sum_{\vect{r}^\dprime_i \in V}\vect{G}_0(\vect{r}^\dprime_j,\vect{r}^\dprime_i)\cdot \vect{E}(\vect{r}^\dprime_i),
\label{eq:form5}
\end{align}
which describes a system of $3N_V$ linear equations for three-dimensional problems, where $N_V$ denotes the number of pixels in the flipped region $V$.

Eq.~(\ref{eq:form5}) can be expressed in matrix form as:
\begin{align}
\left[\mathds{1}_{3N_V} - b \vect{G}_{0,V} \right]\cdot \vect{E}_V = \vect{E}_{0,V},
\label{eq:form6}
\end{align}
% \begin{align}
% \big\{\mathds{1}_{3N_V} - b [G_{0,V}] \big\} \{E_V\} = \{E_{0,V}\},
% \label{eq:form6}
% \end{align}
where we have defined the length $3N_V$ vector as:
\begin{align}
\vect{E}_{V} = \begin{bmatrix} \vect{E}(\vect{r}^\dprime_1) \\ \vdots \\ \vect{E}(\vect{r}^\dprime_{N_V}) \end{bmatrix},
\label{eq:form7}
\end{align}
and $\vect{E}_{0,V}$ is analogously defined for the fields in the unmodified structure. $\mathds{1}_{3N_V}$ is the $3N_V \times 3N_V$ identity matrix. Finally, $\vect{G}_{0,V}$ is a $3N_V \times 3N_V$ matrix consisting of block matrices $\vect{G}_0$ associated with the flipped region $V$:
\begin{align}
\vect{G}_{0,V} = \begin{bmatrix} \vect{G}_0(\vect{r}^\dprime_1,\vect{r}^\dprime_1) & \cdots & \vect{G}_0(\vect{r}^\dprime_1,\vect{r}^\dprime_{N_V}) \\
\vdots & \ddots & \vdots \\
\vect{G}_0(\vect{r}^\dprime_{N_V},\vect{r}^\dprime_1) & \cdots & \vect{G}_0(\vect{r}^\dprime_{N_V},\vect{r}^\dprime_{N_V})
\end{bmatrix}.
\label{eq:form8}
\end{align}
Therefore, we can solve for the resulting field at all flipped pixel locations $\vect{r}^\dprime_i$ in $V$ via Eq.~(\ref{eq:form6}) upon inverting the matrix $\mathds{1}_{3N_V} - b \vect{G}_{0,V}$:
\begin{align}
\vect{E}_V = \left[\mathds{1}_{3N_V} - b \vect{G}_{0,V} \right]^{-1}\cdot \vect{E}_{0,V}.
\label{eq:form9}
\end{align}
We now have all the quantities needed on the right hand side of Eq.~(\ref{eq:form4}) for evaluating the resulting field $\vect{E}(\vect{r})$ at any position $\vect{r} \in \mathcal{D} \cup \{\vect{r}_{m,i}\} $, including, most importantly, the set of monitor positions $\{\vect{r}_{m,i}\}$. The change in FOM that results from the proposed structure modification can in turn be evaluated by Eq.~(\ref{eq:form1}). This completes the first phase of an optimization step where the ``rewards'' (i.e.  changes in the FOM) for arbitrary modifications to the structure are evaluated. The information can then be used to select an optimal modification in the graph traversal. 

Before proceeding further to update the field and Green's function once a modification is chosen, we comment briefly on relevant shapes and sizes for region $V$. There are a number of benefits for considering proposed flipped regions that consist of multiple pixel locations. For example, if there is a minimum feature size that patterned structures must obey due to fabrication constraints, then one way to impose such constraints is by only considering modified regions that satisfy the minimum feature size. Along similar lines, fabricability may require that etched patterns (e.g. in a lithographic process) have a specific or minimum depth. This can also be imposed by considering ``pillars'' of modified material that are of a given height.

\subsection{Updating fields and Green's functions} \label{subsec:updates}

Once a particular structural modification is selected, $\vect{E}_0(\vect{r})$ and $\vect{G}_0(\vect{r},\vect{r}^\prime)$ must be updated before starting the next optimization step. 
In the following discussion, we shall use superscripts ``old'' and ``new'' to denote the $\vect{E}_0$ and $\vect{G}_0$ quantities before and after updating. The updated field $\vect{E}^\textrm{new}_0(\vect{r})$ can be obtained by exactly following the equations in Sec.~\ref{subsec:deltafom}, with the substitutions $\vect{E}_0\rightarrow\vect{E}^\mathrm{old}_0$, $\vect{E}\rightarrow\vect{E}^\mathrm{new}_0$, and $\vect{G}_0\rightarrow\vect{G}^\mathrm{old}_0$. The region $V$ and its constituent pixel locations $\vect{r}^\dprime_i$ should also be reinterpreted as a \textit{selected} modified region rather than a proposed modified region.

The update equations for the Green's function are derived in a similar manner as those for the field. The starting point is now Eq.~(\ref{eq:form2}), which after discretizing and substituting $\vect{G}_0\rightarrow\vect{G}^\mathrm{old}_0$ and $\vect{G}\rightarrow\vect{G}^\mathrm{new}_0$, becomes:
\begin{align}
\vect{G}_0^{\mathrm{new}}(\vect{r},\vect{r}^\prime) =&\: \vect{G}_0^{\mathrm{old}}(\vect{r},\vect{r}^\prime) \nonumber \\&+ b\sum_{\vect{r}^\dprime_i \in V}\vect{G}_0^\mathrm{old}(\vect{r},\vect{r}^\dprime_i)\cdot \vect{G}_0^\mathrm{new}(\vect{r}^\dprime_i,\vect{r}^\prime).
\label{eq:form10}
\end{align}
We first solve the system of equations for positions $\vect{r}$ equal to the flipped pixels $\vect{r}_j^\dprime$:
\begin{align}
\vect{G}_0^{\mathrm{new}}(\vect{r}_j^\dprime,\vect{r}^\prime) =&\: \vect{G}_0^{\mathrm{old}}(\vect{r}_j^\dprime,\vect{r}^\prime) \nonumber \\ 
&+ b\sum_{\vect{r}^\dprime_i \in V}\vect{G}_0^\mathrm{old}(\vect{r}_j^\dprime,\vect{r}^\dprime_i)\cdot \vect{G}_0^\mathrm{new}(\vect{r}^\dprime_i,\vect{r}^\prime),
\label{eq:form11}
\end{align}
For a fixed $\vect{r}^\prime$, we then have a matrix equation analogous to Eq.~(\ref{eq:form6}) for the field:
\begin{align}
\left[\mathds{1}_{3N_V} - b \vect{G}_{0,V}^\mathrm{old} \right]\cdot \vect{G}^\mathrm{new}_{0,V}(\vect{r}^\prime) = \vect{G}^\mathrm{old}_{0,V}(\vect{r}^\prime),
\label{eq:form12}
\end{align}
where $\vect{G}_{0,V}^\mathrm{new|old}(\vect{r}^\prime)$ are $3N_V\times 3$ \textit{matrices} defined as:
\begin{align}
\vect{G}_{0,V}^\mathrm{new|old}(\vect{r}^\prime) = \begin{bmatrix} \vect{G}^\mathrm{new|old}_0(\vect{r}_1^\dprime,\vect{r}^\prime) \\
\vdots\\
\vect{G}^\mathrm{new|old}_0(\vect{r}_{N_V}^\dprime,\vect{r}^\prime)
\end{bmatrix}.
\label{eq:form13}
\end{align}
The definition of $\vect{G}^\mathrm{old}_{0,V}$ is given by Eq.~(\ref{eq:form8}). Thus, we have:
\begin{align}
\vect{G}^\mathrm{new}_{0,V}(\vect{r}^\prime) = \left[\mathds{1}_{3N_V} - b \vect{G}^\mathrm{old}_{0,V} \right]^{-1}\cdot \vect{G}^\mathrm{old}_{0,V}(\vect{r}^\prime),
\label{eq:form14}
\end{align}
which can then be used in Eq.~(\ref{eq:form10}) to obtain the updated Green's functions  $\vect{G}^\mathrm{new}_0(\vect{r},\vect{r}^\prime)$ for arbitrary $\vect{r}$ and $\vect{r}^\prime$.

\begin{figure}[t]
    \centering
    \includegraphics[width=\columnwidth]{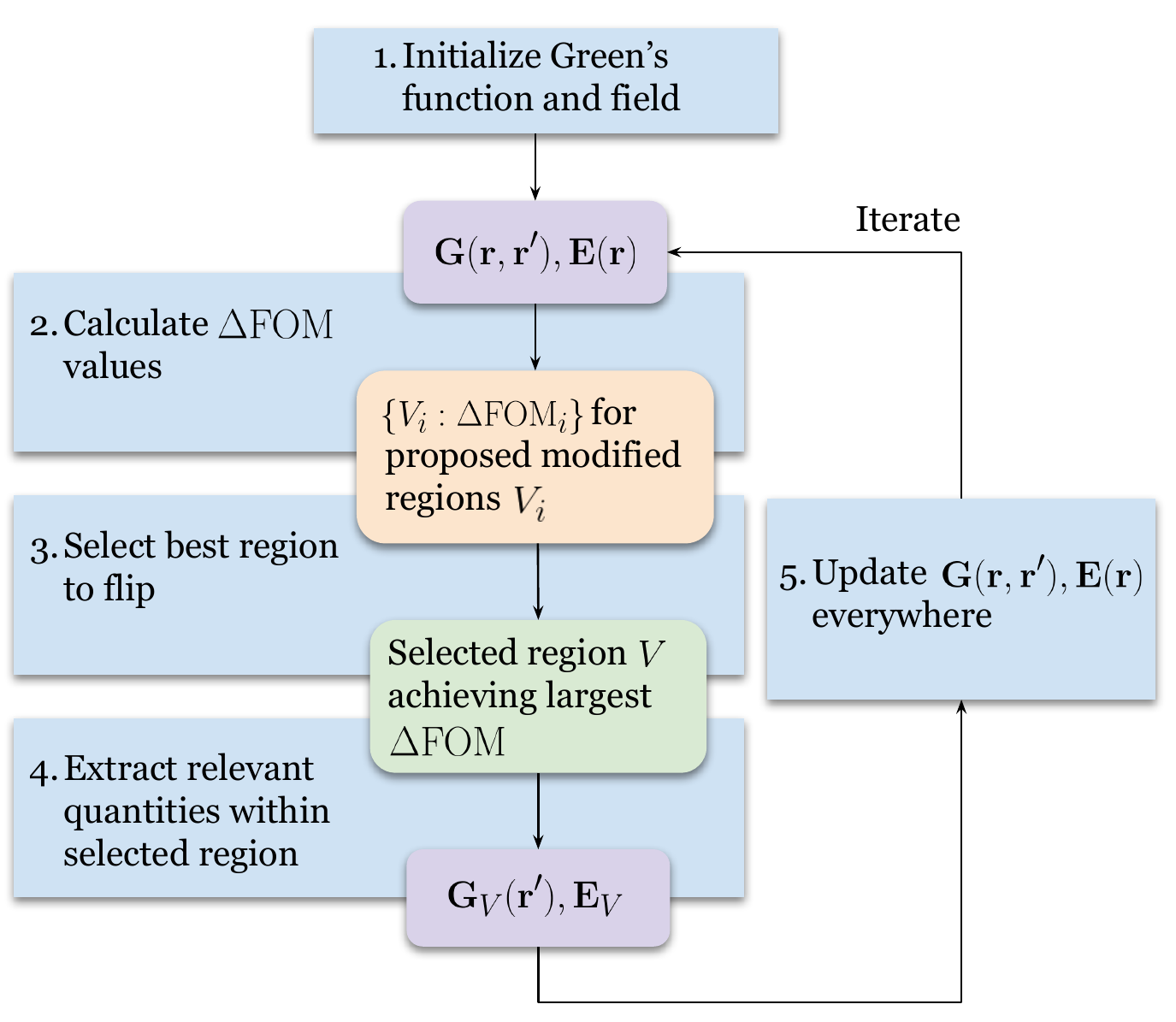}
    \caption{Computational flow chart summarizing the Green's function pixel-by-pixel optimization scheme.}
    \label{fig:form1}
\end{figure}

In summary, each optimization step in the Green's function pixel-by-pixel formalism operates on the assumption that $\vect{E}(\vect{r})$ and $\vect{G}(\vect{r},\vect{r}^\prime)$ are known for all positions within a design region and monitor positions where the FOM is calculated. If these quantities are known, then evaluating the changes in FOM that result from structure modifications in the design region and updating these quantities once a modification is chosen require only the inversion of small matrices ($3N_V\times3N_V$, i.e. of the size of each modification) and tensor multiplications. Both are computationally inexpensive operations. The algorithm is summarized as a flow chart in Fig.~\ref{fig:form1}.

The computational complexity of this algorithm can then be thought of as being dominated by: (a) the computation of the Green's function for an initial structure, and (b) the $\mathcal{O}(N^2)$ memory for storing the two-point Green's function. In principle, (a) can be considered as a computationally intensive but one-off task. Furthermore, for certain classes of initial structures~\cite{paulus:20,Martin2:98}, analytical solutions may exist that eliminate or greatly reduce the complexity of calculating the initial Green's function. For these reasons, (a) is by no means a fundamental barrier to the practical application of this algorithm.

In contrast, the $\mathcal{O}(N^2)$ memory scaling of (b) is the major bottleneck of this algorithm. For even moderately sized design problems with tens of thousands of design pixels, the memory requirement explodes to several hundreds of gigabytes. The natural way to overcome such a severe memory scaling is to parallelize the algorithm and distribute the storage of the Green's function across multiple central processing units (CPU). In the next section, we will show that this is indeed possible with parallel data processing frameworks.

\section{Parallelization} \label{sec:parallelization}

Consider an optimization problem where the design region $\mathcal{D}$ consists of $N=n_x n_y n_z$ pixels, where $n_x$ and $n_y$ are the in-plane sizes, and $n_z$ gives the out-of-plane thickness. Note that we distinguish between sizes in the in-plane and out-of-plane directions because the design region in many applications has a thickness much smaller than the sizes of its lateral dimensions, i.e. $n_z \ll n_x,n_y$. One example is that of integrated photonic devices where the etch depth~\cite{Cheben1:18,Piggott1:15} is typically much smaller than the design region. Another class of examples are metalenses~\cite{Khorasaninejad:16,Akidau1:15,Cheben1:18,Chung1:20} for which the thickness of the lens is also much smaller than the diameter. For clarity of presentation, we will also restrict our attention to the case of a single monitor position $\vect{r}_m$ which we assume to lie outside of the design region. It is trivial to extend to the case of multiple monitor positions.

The quantities that must be stored are $\vect{G}(\vect{r},\vect{r}^\prime)$ and $\vect{E}(\vect{r})$. Note however that the problematic $\mathcal{O}(N^2)$ memory complexity in the algorithm originates only from the part of $\vect{G}(\vect{r},\vect{r}^\prime)$ corresponding to the pairs of positions $\vect{r}$ and $\vect{r}^\prime$ within the design region. It is therefore convenient for accounting purposes to separate the \textit{design region only} Green's function $\vect{G}_{dr}(\vect{r},\vect{r}^\prime)$, where $\vect{r},\vect{r}^\prime\in\mathcal{D}$, from the \textit{monitor point} Green's function $\vect{G}_m(\vect{r})\equiv \vect{G}(\vect{r}_m,\vect{r})$ for $\vect{r}\in\mathcal{D}$. For consistency, we also separate the design region field $\vect{E}_{dr}(\vect{r})$, where $\vect{r}\in\mathcal{D}$, from the single field value at the monitor position $\vect{E}_m \equiv \vect{E}(\vect{r}_m)$.

\begin{figure}[t]
    \centering
    \includegraphics[width=\columnwidth]{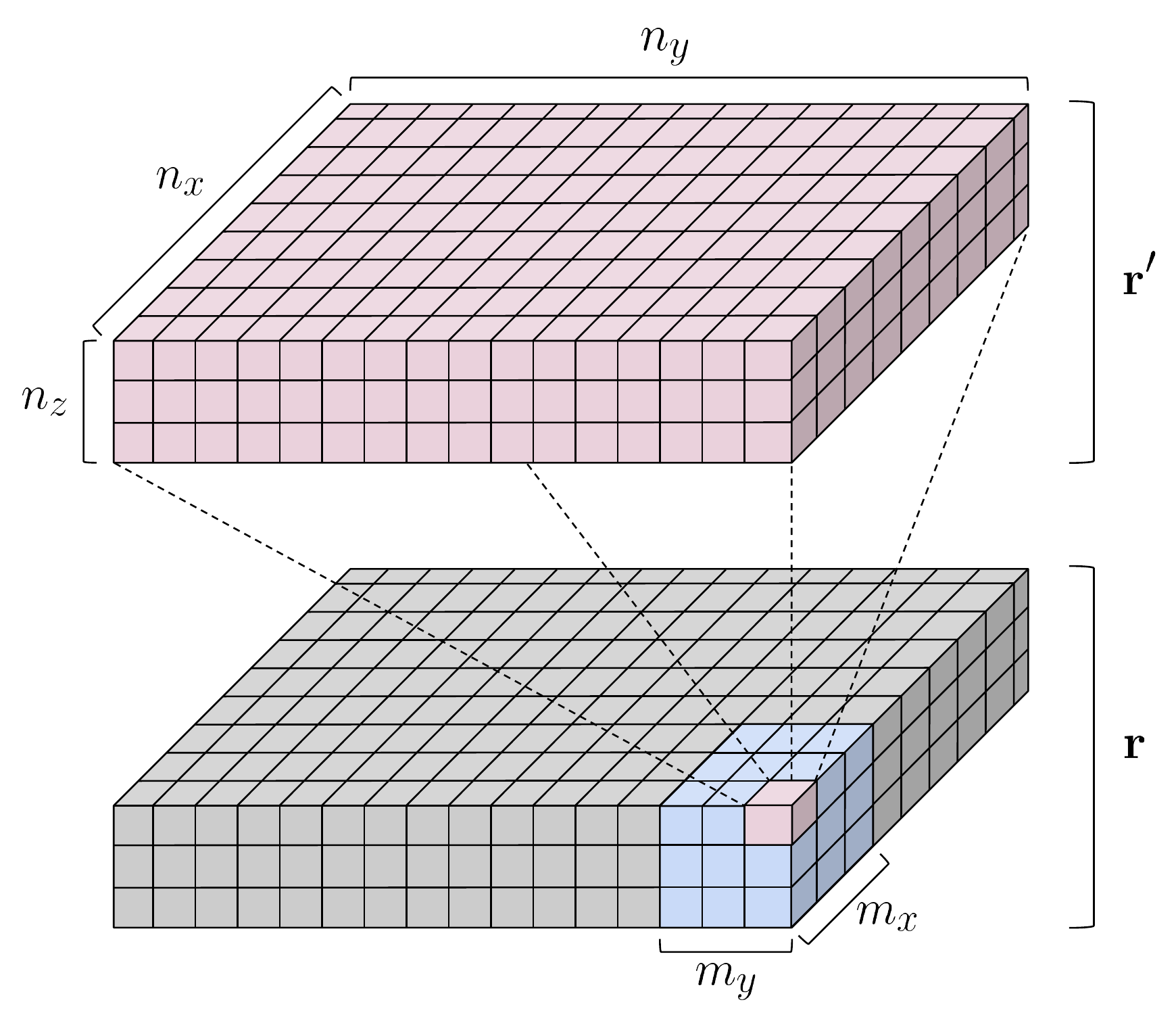}
    \caption{Depiction of how the design region Green's function $\vect{G}_{dr}(\vect{r},\vect{r}^\prime)$ is chunked in the spaces of the two position variables in the distributed implementation. Our strategy is to only split the space of the first position variable $\vect{r}$ into chunks of size $M=m_x m_y n_z \ll N$ while keeping the full size $N$ in the space of the second position variable $\vect{r}^\prime$. This drastically reduces the memory scaling \textit{of each chunk} to $\mathcal{O}(MN)$ such that it can fit into the RAM of an individual CPU worker for considered applications.}
    \label{fig:par1}
\end{figure}

Our first task is to partition $\vect{G}_{dr}(\vect{r},\vect{r}^\prime)$ into divisions that fit into the random-access memory (RAM) of a single worker (note: the amount of RAM available per worker is specific to the computational platform used). Because there are two position variables involved, there are potentially many valid partitioning schemes. Our strategy is to chunk the space of the first position variable $\vect{r}$ while keeping the full size of the second position variable $\vect{r}^\prime$. Picking this particular strategy keeps the implementation relatively simple while the chunk memory requirements remain tractable for the problems considered in this paper. Arbitrary (more granular) chunking strategies can be used in the case of bigger problems or when the RAM available per worker is scarce. Our strategy is illustrated in Fig.~\ref{fig:par1} in which the top and bottom blocks, each of size $N$ pixels, respectively represent the spaces of the two position variables. The direct product of these two spaces then represents all position combinations of the full $\vect{G}_{dr}(\vect{r},\vect{r}^\prime)$ with $\mathcal{O}(N^2)$ elements. The colored region represents a single chunk in our strategy. Namely, the space of $\vect{r}$ is split into chunks of size $M=m_x m_y n_z$ with $m_x \ll n_x$ and $m_y \ll n_y$. We denote the $k^\mathrm{th}$ chunk of $\vect{G}_{dr}(\vect{r},\vect{r}^\prime)$ as $\vect{G}_{dr}^{(k)}$. The memory scaling of $\vect{G}_{dr}^{(k)}$ is then reduced to $\mathcal{O}(MN)$, where $M$ can be $\mathcal{O}(1)$ and, in the extreme case, $M = 1$. Note that we did not split in the out-of-plane direction since we assume $n_z$ is already small. This choice will be particularly convenient for the demonstration in Sec.~\ref{sec:metalens} for performing multipixel ``pillar updates'' where each proposed structure modification flips pixels through the thickness of the design region. Finally, the quantities $\vect{G}_m(\vect{r})$ and $\vect{E}_{dr}(\vect{r})$ do not necessarily need to be chunked because they are only functions of one position and hence already have $\mathcal{O}(N)$ space requirement. Nevertheless, for consistency, we will also split them into chunks of size $M$, denoted as $\vect{G}_m^{(k)}$ and $\vect{E}_{dr}^{(k)}$ respectively. We therefore define the $k^\mathrm{th}$ partition to consist of $\{\vect{G}_{dr}^{(k)},\vect{G}_m^{(k)}, \vect{E}_{dr}^{(k)}\}$.

Now that we have defined the composition of each partition, we must consider how the main computations in the Green's function pixel-by-pixel optimization scheme can be performed within each partition. Recall from Sec.~\ref{sec:formalism} that each optimization step consists of two main phases: the calculation of $\Delta \mathrm{FOM}$ for various proposed structure modifications, and the updates to the Green's functions and fields after a modification is chosen. 

Let proposed modification $V$ be contained within partition $k$. For the $\Delta \mathrm{FOM}$ calculation, the relevant equations are Eqs.~(\ref{eq:form4})-(\ref{eq:form9}). In particular, the inputs needed for determining all resulting $\vect{E}(\vect{r}_i^\dprime)$ used in Eq.~(\ref{eq:form4}) are the sets of all $\vect{G}_0(\vect{r}_i^\dprime,\vect{r}_j^\dprime)$ and $\vect{E}_0(\vect{r}_i^\dprime)$. Because all $\vect{r}^\dprime$ belong to $V$, and $V$ exists within the design region partition $k$, we have all required inputs contained within the chunked quantities $\vect{G}_{0,dr}^{(k)}$ and $\vect{E}_{0,dr}^{(k)}$. We now return to Eq.~(\ref{eq:form4}). Evaluating it at the monitor position $\vect{r}=\vect{r}_m$ and reinterpreting quantities with respect to a partition $k$ (e.g. $\vect{G}_0(\vect{r}_m,\vect{r}_i^\dprime)=\vect{G}_{0,m}^{(k)}(\vect{r}^\dprime_i)$), we obtain:
\begin{align}
\vect{E}_m &= \vect{E}_{0,m} + b\sum_{\vect{r}^\dprime_i \in V}\vect{G}_{0,m}^{(k)}(\vect{r}^\dprime_i)\cdot \vect{E}_{dr}^{(k)}(\vect{r}^\dprime_i),
\label{eq:para1}
\end{align}
which can finally be used to evaluate $\Delta \mathrm{FOM}$ for the proposed modification. We conclude that for any proposed modified region contained within a partition, the partition possesses all the information needed for computation of the $\Delta \mathrm{FOM}$ without the need for any data to be communicated from other partitions.

We can hence imagine that during the first phase of an optimization step in the parallelized implementation, each partition is assigned to a worker. Given partition $k$, the task of the worker is to independently compute the $\Delta\mathrm{FOM}$ values that result from proposed modified regions $V_{k\ell}$ within the partition, where $\ell$ indexes the proposed modified regions within partition $k$. Each worker then emits pairs consisting of the modification proposal $V_{k\ell}$ and corresponding $\Delta\mathrm{FOM}$. These pairs are then aggregated over all the partitions to a single proposal through a reduction operation. Each reduction operation is given a set of partitions $K$ of $\{V_{k\ell}\}^{k \in K}$ and decides which modification $V_{k\ell}$ to select based on the $\Delta\mathrm{FOM}$ values. It emits the chosen modification and the process continues until a single modification $V^*$ originating from partition $k^*$ remains. This concludes the first phase of the optimization step.

Next, we turn to the update step after the single region $V^*$ originating from partition $k^*$ is chosen and flipped. From the discussion in Sec.~\ref{subsec:updates}, we must first compute the updated quantities at the flipped pixel positions:
\begin{align}
\vect{G}^{\mathrm{new}}_{0,{V^*}}(\vect{\vect{r}^\prime}) &= \begin{bmatrix}
\vect{G}^\mathrm{new}_0(\vect{r}_1^\dprime,\vect{r}^\prime) \\
\vdots\\
\vect{G}^\mathrm{new}_0(\vect{r}_{N_{V^*}}^\dprime,\vect{r}^\prime)
\end{bmatrix}\\
\vect{E}^{\mathrm{new}}_{0,{V^*}} &= 
\begin{bmatrix}
\vect{E}^\mathrm{new}_0(\vect{r}_1^\dprime) \\
\vdots\\
\vect{E}^\mathrm{new}_0(\vect{r}_{N_{V^*}}^\dprime)
\end{bmatrix},
\label{eq:para2}
\end{align}
The information needed to compute these quantities is available only in the partition $k^*$. Therefore, partition $k^*$ must communicate these quantities to all other partitions. Its data are fetched through a filter operation and then broadcast. The quantities $\{\vect{G}_{dr}^{(k)},\vect{G}_m^{(k)},
\vect{E}_{dr}^{(k)}\}$ in each partition $k$ can then be independently updated via Eqs.~(\ref{eq:form4}) and~(\ref{eq:form10}).

\begin{figure*}[ht]
    \centering
    \includegraphics[width=\textwidth]{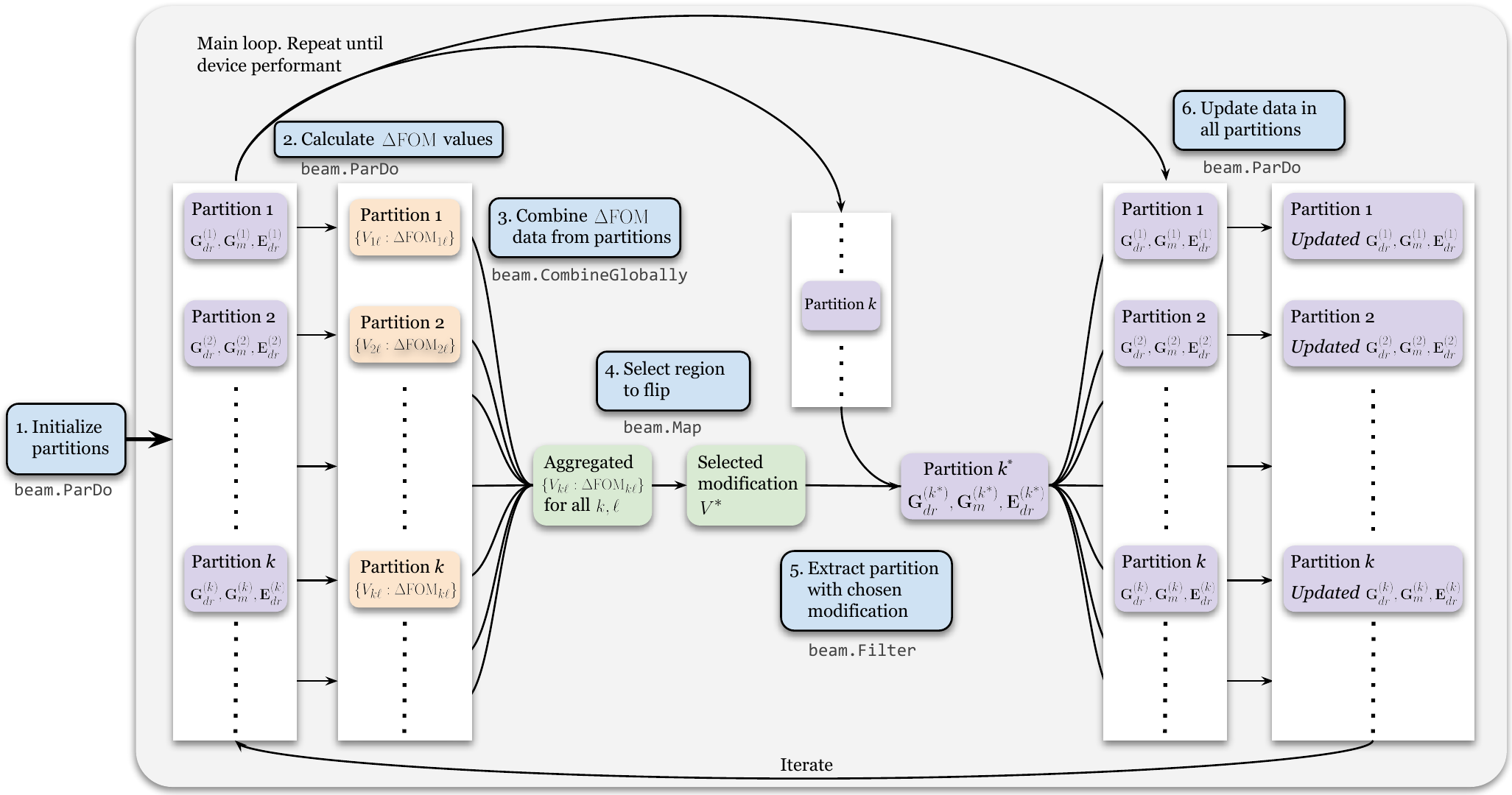}
    \caption{Flow diagram representing our parallelized implementation of the Green's function pixel-by-pixel optimization scheme as an Apache Beam data processing pipeline. The $\Delta\mathrm{FOM}$ calculation and partition update stages are implemented as \texttt{beam.ParDo} operations. The $\Delta\mathrm{FOM}$ data aggregation stage is implemented as a \texttt{beam.CombineGlobally} operation. The identification of the partition $k^*$ containing the selected modification region $V^*$ is implemented as a \texttt{beam.Filter} operation. Note that the initialization of the partitions can take many forms. For example, pre-computed Green's function and field values may be read from disk. Alternatively, if analytical expressions exist, then each partition may initialize its own Green's functions and fields.}
    \label{fig:par2}
\end{figure*}

To summarize the discussion so far, we have proposed a parallelization scheme for the Green's function based pixel-by-pixel optimization algorithm in which the Green's function---the leading source of memory complexity---is partitioned into chunks that effectively scale as $\mathcal{O}(N)$ as opposed to $\mathcal{O}(N^2)$. Under this partitioning convention, the two main computational phases of an optimization step can be executed across all the partitions \textit{independently} without the need for any inter-partition communication. The stage that does require communication is when the $\Delta \mathrm{FOM}$ values for the proposed modifications are aggregated across all partitions. Viewed through the lens of parallel programming, this step is a \textit{reduction operation} over the partitions, and efficient implementations for such operations are readily available in various large scale data processing models.

The parallelized optimization scheme described above can thus be implemented as a large scale data processing pipeline. In this work we choose the Apache Beam~\cite{Akidau1:15,Apache:21} programming model, which is an open source library for data-parallel processing pipelines. It is highly portable and integrated with distributed processing backends like Apache Spark or Google Cloud Dataflow~\cite{DataflowGCP} which in turn provide many features like autoscaling of resources, monitoring, data parallelism and fault tolerance mechanisms, among others. A flow diagram of our Apache Beam pipeline is shown in Fig.~\ref{fig:par2}, color coded for the main steps to align with the steps of the general algorithm from Fig.~\ref{fig:form1}. The $\Delta\mathrm{FOM}$ calculation and partition update stages are implemented as \texttt{beam.ParDo} (parallel map) operations while the $\Delta\mathrm{FOM}$ data aggregation stage is implemented as a \texttt{beam.CombineGlobally} (reduce) operation. In the next section, we will exercise our massively parallel implementation to design a high NA focusing metalens with a design region size that would be out of reach without incorporating the data parallelism.

\section{Demonstration on metalens optimization} \label{sec:metalens}

\begin{figure}[t]
    \centering
    \includegraphics[width=0.9\columnwidth]{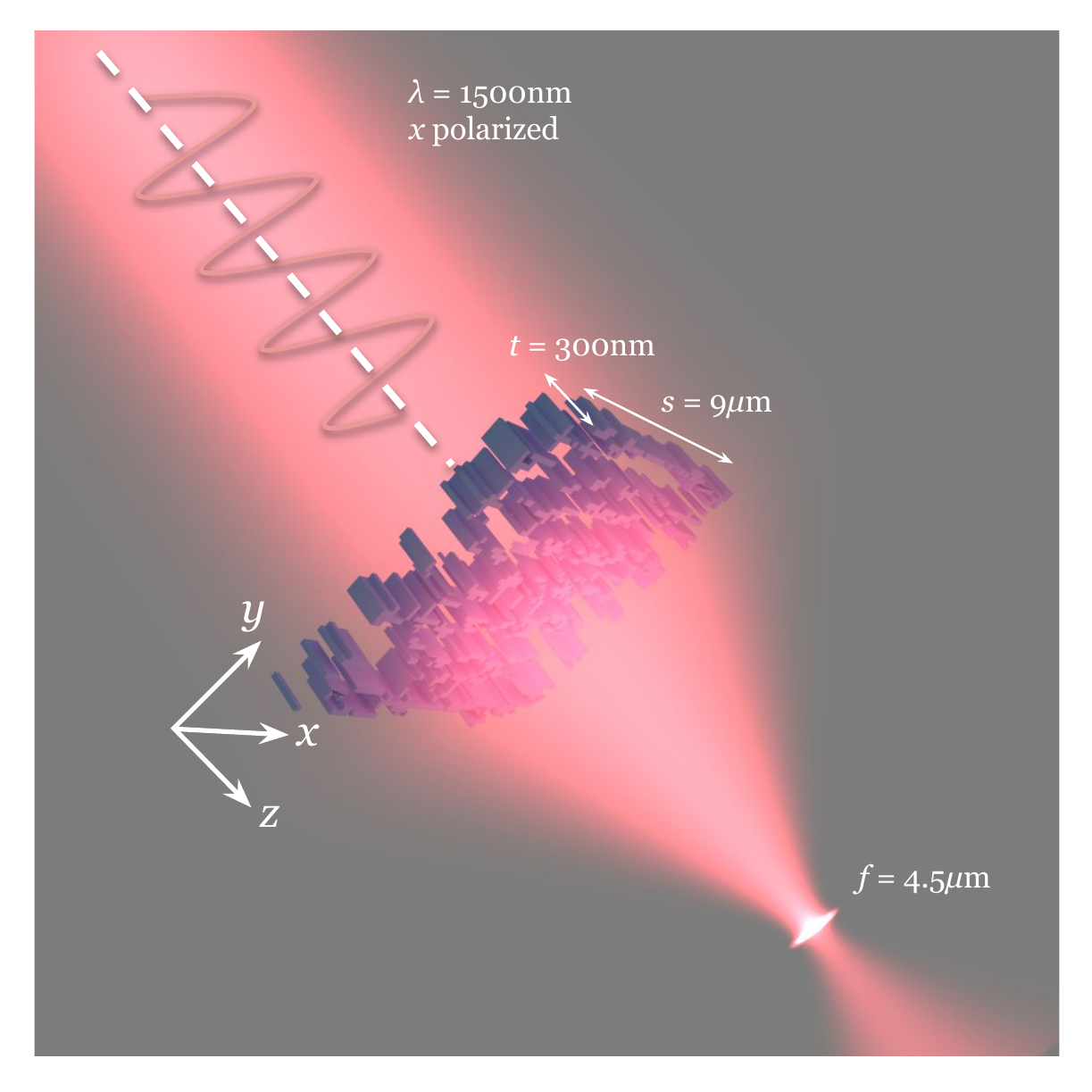}
    \caption{An illustration of the configuration of the high NA focusing metalens optimized using our massively parallel Green's function based pixel-by-pixel optimization scheme. The metalens design region consists of a square region with a side length of $s = 9\mu\textrm{m}$ and a thickness of $t = 300\textrm{nm}$ surrounded by vacuum. The goal of the optimization is to pattern the design region with silicon such that the resulting metalens focuses an $x$-polarized plane wave light source with wavelength $\lambda=1500\textrm{nm}$. The targeted focal length is $f = 4.5 \mu\textrm{m}$, corresponding to a NA of 0.7.}
    \label{fig:results1}
\end{figure}

We illustrate the scale of problems that can be tackled with our massively parallel scheme by optimizing a high NA focusing metalens. The configuration is illustrated in Fig.~\ref{fig:results1}. Our design region consists of a square slab with a side length of $9\mu\textrm{m}$ and a thickness of $300\textrm{nm}$. We take the initial background system to consist entirely of vacuum ($\varepsilon_0=1$) such that the initial Green's function is known analytically~\cite{Martin2:98}. An $x$-polarized plane wave of wavelength $\lambda=1500\textrm{nm}$ is incident normally, and the optimization objective is to maximize the intensity of the diffracted light at a focal point located $4.5\mu\textrm{m}$ ($\textrm{NA} = 0.7$) from the design region by patterning the design region with blocks of silicon ($\varepsilon=12$) that extend through the thickness of the lens.

Although our primary objective here is to demonstrate the ability of our scheme to scale to large problem sizes, we note that the problem configuration described above might model an \textit{in-fiber focusing metalens} in which the facet of an optical fiber is patterned to focus the outgoing light~\cite{Yang1:19,Kim1:20,Principe1:17,Asadollahbaik1:20}. For this particular problem, the most popular optimization method used in the literature is the phase profile matching approach where a slowly varying phase change distribution analytically known to effect focusing behavior is modeled by stitching together large, predefined unit cell structures that have only a small number of degrees of freedom---e.g. nanorods several hundreds of nanometers in length, each configured by an orientation angle. This approach has proven to be extremely efficient in the design of large area metasurfaces~\cite{Phan1:19}. However, for metalenses with high NA and therefore fast varying phase changes, the phase profile stitching approach breaks down~\cite{Lin:2019}. In contrast, the Green's function approach is naturally free of such constraints. Furthermore, by optimizing directly in the pixel representation rather than with large unit cells, the Green's function approach allows for a much larger number of design degrees of freedom per unit area, significantly expanding the landscape of potential designs.

\begin{figure}[t!]
    \centering
    \includegraphics[width=\columnwidth]{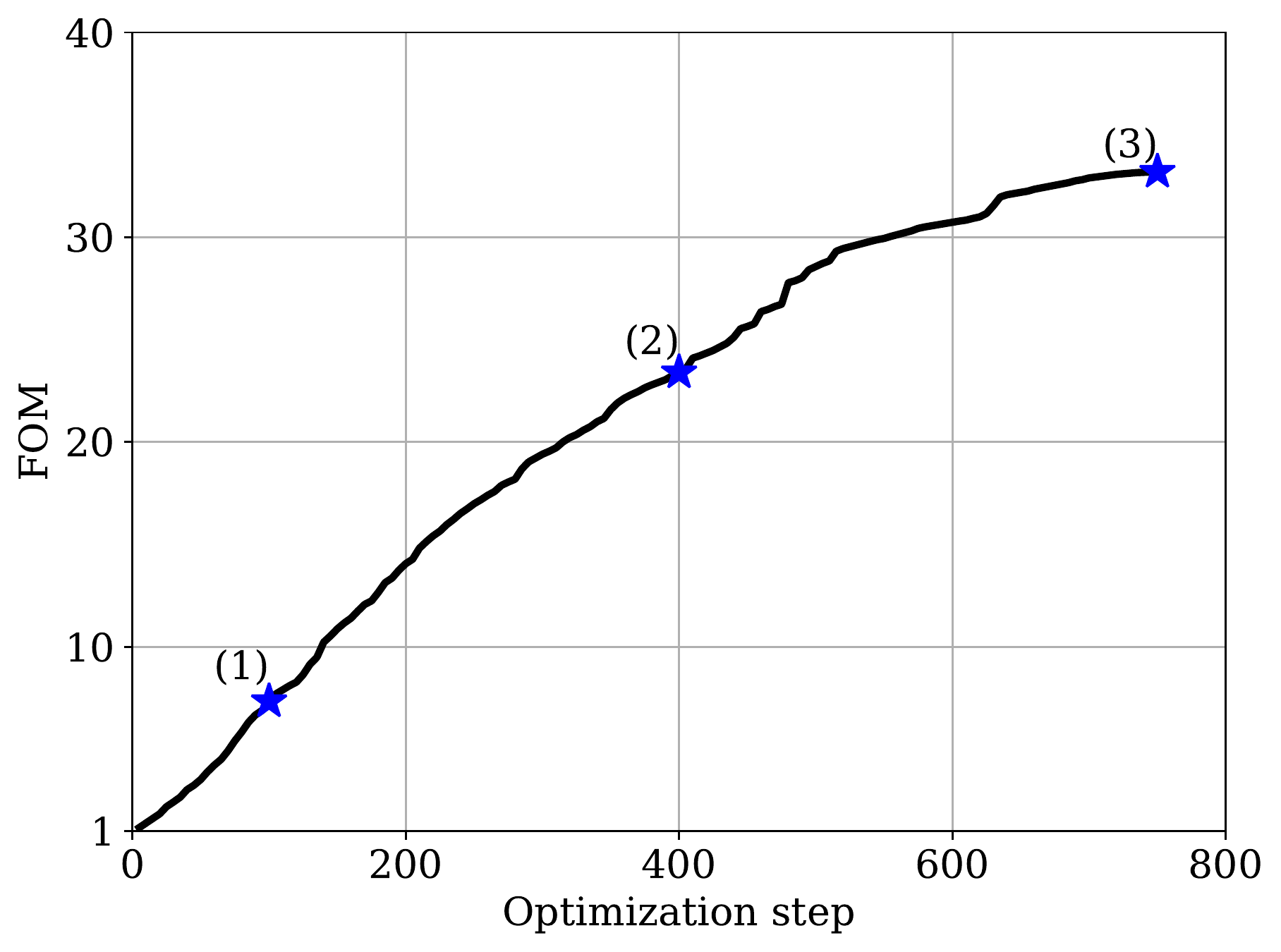}
    \caption{FOM versus optimization step for an optimization of a focusing metalens as described in the main text. The FOM is defined to be the ratio of the achieved electric field intensity at the focal point relative to the initial intensity. The three points (at steps 100, 400, and 750) labeled by star symbols along the trajectory correspond to three designs studied in detail in Fig.~\ref{fig:results3}.}
    \label{fig:results2}
\end{figure}

In Fig.~\ref{fig:results2}, we plot the evolution of the FOM over the course of an optimization of the metalens system. The FOM is defined to be the ratio of the achieved electric field intensity at the focal point relative to the initial intensity. A resolution of $75\textrm{nm}$ per pixel was used for the discretization of the Green's function and electric field, such that the Green's function in the $9 \mu\textrm{m}\times9\mu\textrm{m}\times 300 \textrm{nm}$ design region consisted of $(120\times120\times 4)^2$ complex valued numbers. A lateral minimum feature size of $150\textrm{nm}$ was imposed by only considering modified regions $V_{k\ell}$ that are $2\times2$ squares in the lateral dimensions of the design region. As expected from a fully greedy search strategy, the FOM monotonically increases throughout the optimization until step 750 (labeled point (3) in Fig.~\ref{fig:results2}) when no blocks that would increase FOM can be found.

Computationally, the partitioning of the design space was implemented by chunking the $120\times120$ pixel sized lateral design region into $K=60\times60=3600$ partitions, each of size $2\times2$---i.e. the size of each partition equals the minimum feature size imposed. The computation of the pipeline was distributed across a fixed pool of 2000 workers (each utilizing at most 1 CPU and $1\textrm{GiB}$ of RAM). The sum of the sizes of the inputs taken over all partitions in each optimization step was on average $220\textrm{GiB}$, and each optimization step took on average 3.7 minutes. We note that our optimization scheme also works at higher resolutions, for example at $50\textrm{nm}$ per pixel, where the memory requirement for the inputs at each step increases to $\sim 2\textrm{TiB}$, in accordance with the $\mathcal{O}(N^2)$ scaling.

\begin{figure*}[ht]
    \centering
    \includegraphics[width=0.8\textwidth]{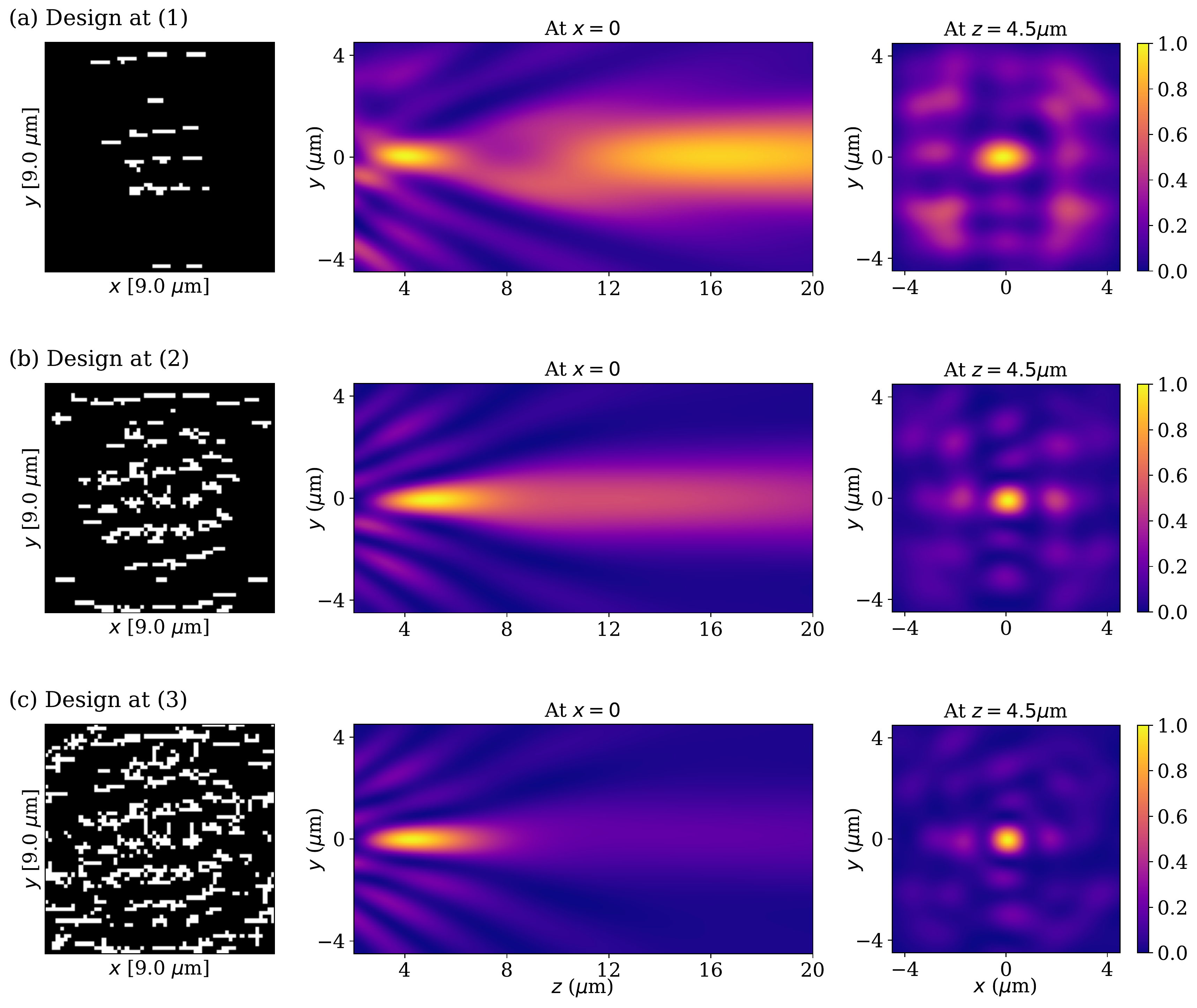}
    \caption{The designs and FDTD validated relative field intensity distributions for (a) step 100, (b) step 400, and (c) step 750 along the optimization trajectory from Fig.~\ref{fig:results2} (indicated by the locations of the three star symbols, (1), (2), and (3)). For the designs, white represents the locations of patterned silicon pillars, while black represents the background vacuum. The targeted focal length is $4.5\mu\textrm{m}$, which corresponds to $\textrm{NA}=0.7$ for a metalens with side length $9\mu\textrm{m}$. Note that a minimum feature size of $150\textrm{nm}$ is imposed by considering modified regions that are $2\times2$ pixels large in the lateral design region dimensions. For the relative intensity distribution plots, the middle panels show the $y$-$z$ plane cross section, while the right panels show the $x$-$y$ focal plane cross section at $z=4.5 \mu\textrm{m}$. The metalens is located at $z=0$. As desired, the relative intensity at the focal point increases substantially as the optimization progresses.}
    \label{fig:results3}
\end{figure*}

The left panels in Fig.~\ref{fig:results3} respectively show the metalens design at steps 100, 400, and 750 of the optimization (labeled by the star symbols in Fig.~\ref{fig:results2}). Qualitatively, the optimizer tends to place slabs of silicon material oriented along the $x$ axis. This aligns with physical intuition as the incident plane wave is polarized along $x$. Larger currents contributed by the polarization, and therefore larger phase changes, can then be induced when the dielectric patterns are oriented along the polarization direction. Validation simulations of the three designs were performed with the finite-difference time-domain (FDTD) method~\cite{Taflove1:05}, using the open-source software package Meep~\cite{Oskooi1:10}. The middle and right panels in Fig.~\ref{fig:results3} show the relative intensity distributions of the focal spot: the middle panels show the $y$-$z$ plane cross section, while the right panels show the $x$-$y$ plane cross section at $z=4.5 \mu\textrm{m}$ (the focal plane). The intensity distribution is initially quite diffuse at step 100, but becomes increasingly concentrated at the focal point by the end of the optimization at step 750, demonstrating that the optimizer is indeed driving the design towards the objective.

\section{Discussions and Conclusions} \label{sec:summary}

We have already enumerated several inherent advantages of pixel-by-pixel topology optimization, namely, the circumvention of full-wave simulations, binarization by construction of the scheme, and the ease of enforcing minimum feature size constraints. We highlight another feature of the pixel-by-pixel paradigm: it allows for the reframing of the nanophotonic topology optimization problem as a graph traversal, where the optimizer can be considered as an \textit{agent} selecting among a discrete set of \textit{actions} (i.e. which pixels to flip) at each optimization step. In other words, the topology optimization can now be naturally reinterpreted as an RL problem.

The policy of the agent utilized in the results presented in Section~\ref{sec:metalens} is the simple greedy policy where the maximum positive $\Delta\textrm{FOM}$ inducing action is always selected. However, completely greedy policies do not, in general, lead to the best results, especially for loss landscapes as complicated as those in photonic inverse design problems. In the language of RL, effective agents are those that strike the right balance between \textit{exploration} and \textit{exploitation}. 

\begin{figure}[t]
    \centering
    \includegraphics[width=\columnwidth]{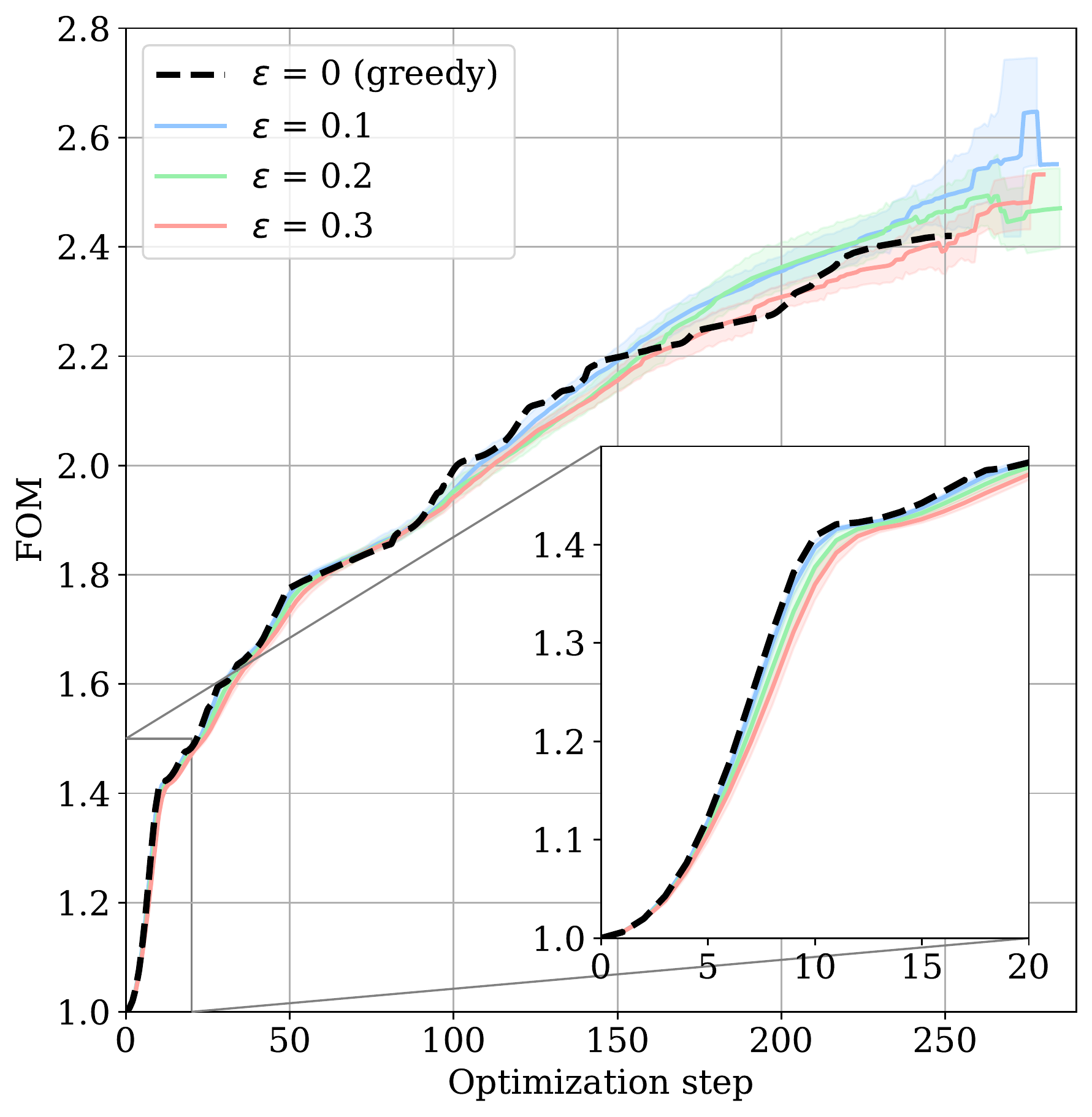}
    \caption{Comparison of greedy and $\epsilon$-greedy optimization trajectories for the optimization of a small $21\times21$ pixel sized focusing metalens targeting a NA of 0.7. For the $\epsilon$-greedy trajectories, the lines represent the mean over 20 independent runs for each value of $\epsilon$; the colored regions represent the range of two standard deviations of the FOM values achieved. All optimizations are run until no more actions that increase the FOM remain. The inset shows the data magnified near the start of the optimizations, showing the greedy strategy (black) outperforming the non-greedy strategies. The advantage does not persist, however, with $\epsilon=0.1,0.2$ cases outperforming the greedy strategy when optimizations are run to termination.}
    \label{fig:disc1}
\end{figure}

To showcase this concept, we performed optimizations of a small scale $21\times21$ pixel sized focusing metalens at a resolution of $75\textrm{nm}$ per pixel targeting a $\textrm{NA}$ of 0.7, using the $\epsilon$-greedy policy. The $\epsilon$-greedy policy is a simple strategy to introduce an element of exploration to a search algorithm: configured by a parameter $\epsilon\in[0,1]$, the agent acts greedily (exploits) with probability $1-\epsilon$, but randomly selects among a set of locally non-optimal actions (explores) with probability $\epsilon$. In our demonstration, the agent randomly selects among the top four $\textrm{FOM}$ increasing actions when it explores. The results are shown in Fig.~\ref{fig:disc1} where the optimization trajectory for the greedy ($\epsilon=0$, dashed black line) are plotted alongside the mean over 20 trajectories for non-greedy strategies with $\epsilon=0.1,0.2,0.3$ (blue, green, and red lines). The shaded colored regions represent the range of two standard deviations for the non-greedy trajectories. All optimizations are run until no more actions that increase the FOM remain.

Focusing first on the magnified inset of Fig.~\ref{fig:disc1}, the greedy strategy initially outperforms all non-greedy strategies. However, it is overtaken by the $\epsilon=0.1, 0.2$ strategies towards the end, which not only find designs that outperform the greedily obtained terminal design by $15\%$ in FOM, but also perform better on average. We expect this advantage to be amplified for larger sized problems involving longer optimization trajectories. For $\epsilon=0.3$, the performance degrades to approximately the level of the greedy strategy, which can be ascribed to an overemphasis on exploration. Our simple example illustrating the effects of including non-greediness hints at the potential of applying techniques from the vast and topical field of RL to the field of nanophotonic inverse design---a connection that becomes all the more apparent and ripe for exploration through the lens of the Green's function based pixel-by-pixel optimization paradigm.

In conclusion, in this paper we have demonstrated a massively parallel implementation of the Green's function based pixel-by-pixel optimization scheme---a scheme that would otherwise be severely limited in the scope of applications by the $\mathcal{O}(N^2)$ memory scaling for the storage of the two-point Green's function. Researchers and designers in industry and academia alike now have access to high performance computing clusters as well as commercial cloud computing platforms. Thus, the method we have proposed elevates the Green's function based optimization paradigm to a new level of scale and practicality comparable to that of adjoint variable, gradient-based approaches. Furthermore, by reinterpreting the photonic inverse design problem as a graph traversal, immediate connections to the field of RL can be made. In combination with our method for scaling the Green's function scheme, we believe that opportunities await for tackling large scale photonic inverse design problems with RL inspired techniques. Future work will focus on further developing this connection, as well as extending the Green's function scheme to optimize for more realistic configurations of the metalens (e.g. inclusion of a substrate layer, imposition of spatial symmetries) and beyond (e.g. integrated photonic devices).


\begin{thebibliography}{34}
\expandafter\ifx\csname natexlab\endcsname\relax\def\natexlab#1{#1}\fi
\expandafter\ifx\csname bibnamefont\endcsname\relax
  \def\bibnamefont#1{#1}\fi
\expandafter\ifx\csname bibfnamefont\endcsname\relax
  \def\bibfnamefont#1{#1}\fi
\expandafter\ifx\csname citenamefont\endcsname\relax
  \def\citenamefont#1{#1}\fi
\expandafter\ifx\csname url\endcsname\relax
  \def\url#1{\texttt{#1}}\fi
\expandafter\ifx\csname urlprefix\endcsname\relax\def\urlprefix{URL }\fi
\providecommand{\bibinfo}[2]{#2}
\providecommand{\eprint}[2][]{\url{#2}}

\bibitem[{\citenamefont{Molesky et~al.}(2018)\citenamefont{Molesky, Lin,
  Piggott, Jin, Vuckovi{\'c}, and Rodriguez}}]{Molesky1:18}
\bibinfo{author}{\bibfnamefont{S.}~\bibnamefont{Molesky}},
  \bibinfo{author}{\bibfnamefont{Z.}~\bibnamefont{Lin}},
  \bibinfo{author}{\bibfnamefont{A.~Y.} \bibnamefont{Piggott}},
  \bibinfo{author}{\bibfnamefont{W.}~\bibnamefont{Jin}},
  \bibinfo{author}{\bibfnamefont{J.}~\bibnamefont{Vuckovi{\'c}}},
  \bibnamefont{and} \bibinfo{author}{\bibfnamefont{A.~W.}
  \bibnamefont{Rodriguez}}, \bibinfo{journal}{Nature Photonics}
  \textbf{\bibinfo{volume}{12}}, \bibinfo{pages}{659} (\bibinfo{year}{2018}).

\bibitem[{\citenamefont{Wang et~al.}(2018)\citenamefont{Wang, Shi, Hughes,
  Zhao, and Fan}}]{Wang1:18}
\bibinfo{author}{\bibfnamefont{J.}~\bibnamefont{Wang}},
  \bibinfo{author}{\bibfnamefont{Y.}~\bibnamefont{Shi}},
  \bibinfo{author}{\bibfnamefont{T.}~\bibnamefont{Hughes}},
  \bibinfo{author}{\bibfnamefont{Z.}~\bibnamefont{Zhao}}, \bibnamefont{and}
  \bibinfo{author}{\bibfnamefont{S.}~\bibnamefont{Fan}}, \bibinfo{journal}{Opt.
  Express} \textbf{\bibinfo{volume}{26}}, \bibinfo{pages}{3236}
  (\bibinfo{year}{2018}),
  \urlprefix\url{http://www.osapublishing.org/oe/abstract.cfm?URI=oe-26-3-3236}.

\bibitem[{\citenamefont{Veronis et~al.}(2004)\citenamefont{Veronis, Dutton, and
  Fan}}]{Veronis1:04}
\bibinfo{author}{\bibfnamefont{G.}~\bibnamefont{Veronis}},
  \bibinfo{author}{\bibfnamefont{R.~W.} \bibnamefont{Dutton}},
  \bibnamefont{and} \bibinfo{author}{\bibfnamefont{S.}~\bibnamefont{Fan}},
  \bibinfo{journal}{Optics letters} \textbf{\bibinfo{volume}{29}},
  \bibinfo{pages}{2288} (\bibinfo{year}{2004}).

\bibitem[{\citenamefont{Hammond et~al.}(2021)\citenamefont{Hammond, Oskooi,
  Johnson, and Ralph}}]{Hammond1:21}
\bibinfo{author}{\bibfnamefont{A.~M.} \bibnamefont{Hammond}},
  \bibinfo{author}{\bibfnamefont{A.}~\bibnamefont{Oskooi}},
  \bibinfo{author}{\bibfnamefont{S.~G.} \bibnamefont{Johnson}},
  \bibnamefont{and} \bibinfo{author}{\bibfnamefont{S.~E.} \bibnamefont{Ralph}},
  \bibinfo{journal}{Opt. Express} \textbf{\bibinfo{volume}{29}},
  \bibinfo{pages}{23916} (\bibinfo{year}{2021}),
  \urlprefix\url{http://www.osapublishing.org/oe/abstract.cfm?URI=oe-29-15-23916}.

\bibitem[{\citenamefont{Sell et~al.}(2017)\citenamefont{Sell, Yang, Doshay,
  Yang, and Fan}}]{Sell1:17}
\bibinfo{author}{\bibfnamefont{D.}~\bibnamefont{Sell}},
  \bibinfo{author}{\bibfnamefont{J.}~\bibnamefont{Yang}},
  \bibinfo{author}{\bibfnamefont{S.}~\bibnamefont{Doshay}},
  \bibinfo{author}{\bibfnamefont{R.}~\bibnamefont{Yang}}, \bibnamefont{and}
  \bibinfo{author}{\bibfnamefont{J.~A.} \bibnamefont{Fan}},
  \bibinfo{journal}{Nano letters} \textbf{\bibinfo{volume}{17}},
  \bibinfo{pages}{3752} (\bibinfo{year}{2017}).

\bibitem[{\citenamefont{Piggott et~al.}(2015)\citenamefont{Piggott, Lu,
  Lagoudakis, Petykiewicz, Babinec, and Vu{\v{c}}kovi{\'c}}}]{Piggott1:15}
\bibinfo{author}{\bibfnamefont{A.~Y.} \bibnamefont{Piggott}},
  \bibinfo{author}{\bibfnamefont{J.}~\bibnamefont{Lu}},
  \bibinfo{author}{\bibfnamefont{K.~G.} \bibnamefont{Lagoudakis}},
  \bibinfo{author}{\bibfnamefont{J.}~\bibnamefont{Petykiewicz}},
  \bibinfo{author}{\bibfnamefont{T.~M.} \bibnamefont{Babinec}},
  \bibnamefont{and}
  \bibinfo{author}{\bibfnamefont{J.}~\bibnamefont{Vu{\v{c}}kovi{\'c}}},
  \bibinfo{journal}{Nature Photonics} \textbf{\bibinfo{volume}{9}},
  \bibinfo{pages}{374} (\bibinfo{year}{2015}).

\bibitem[{\citenamefont{Vercruysse et~al.}(2019)\citenamefont{Vercruysse,
  Sapra, Su, Trivedi, and Vu{\v c}kovi{\'c}}}]{Vercruysse1:19}
\bibinfo{author}{\bibfnamefont{D.}~\bibnamefont{Vercruysse}},
  \bibinfo{author}{\bibfnamefont{N.~V.} \bibnamefont{Sapra}},
  \bibinfo{author}{\bibfnamefont{L.}~\bibnamefont{Su}},
  \bibinfo{author}{\bibfnamefont{R.}~\bibnamefont{Trivedi}}, \bibnamefont{and}
  \bibinfo{author}{\bibfnamefont{J.}~\bibnamefont{Vu{\v c}kovi{\'c}}},
  \bibinfo{journal}{Scientific reports} \textbf{\bibinfo{volume}{9}},
  \bibinfo{pages}{8999} (\bibinfo{year}{2019}).

\bibitem[{\citenamefont{Piggott et~al.}(2017)\citenamefont{Piggott,
  Petykiewicz, Su, and Vu{\v{c}}kovi{\'c}}}]{Piggott2:17}
\bibinfo{author}{\bibfnamefont{A.~Y.} \bibnamefont{Piggott}},
  \bibinfo{author}{\bibfnamefont{J.}~\bibnamefont{Petykiewicz}},
  \bibinfo{author}{\bibfnamefont{L.}~\bibnamefont{Su}}, \bibnamefont{and}
  \bibinfo{author}{\bibfnamefont{J.}~\bibnamefont{Vu{\v{c}}kovi{\'c}}},
  \bibinfo{journal}{Scientific reports} \textbf{\bibinfo{volume}{7}},
  \bibinfo{pages}{1} (\bibinfo{year}{2017}).

\bibitem[{\citenamefont{Schubert et~al.}(2022)\citenamefont{Schubert, Cheung,
  Williamson, Spyra, and Alexander}}]{Schubert1:22}
\bibinfo{author}{\bibfnamefont{M.~F.} \bibnamefont{Schubert}},
  \bibinfo{author}{\bibfnamefont{A.~K.~C.} \bibnamefont{Cheung}},
  \bibinfo{author}{\bibfnamefont{I.~A.~D.} \bibnamefont{Williamson}},
  \bibinfo{author}{\bibfnamefont{A.}~\bibnamefont{Spyra}}, \bibnamefont{and}
  \bibinfo{author}{\bibfnamefont{D.~H.} \bibnamefont{Alexander}},
  \emph{\bibinfo{title}{Inverse design of photonic devices with strict foundry
  fabrication constraints}} (\bibinfo{year}{2022}), \eprint{2201.12965}.

\bibitem[{\citenamefont{Silver et~al.}(2018)\citenamefont{Silver, Hubert,
  Schrittwieser, Antonoglou, Lai, Guez, Lanctot, Sifre, Kumaran, Graepel
  et~al.}}]{Silver1:18}
\bibinfo{author}{\bibfnamefont{D.}~\bibnamefont{Silver}},
  \bibinfo{author}{\bibfnamefont{T.}~\bibnamefont{Hubert}},
  \bibinfo{author}{\bibfnamefont{J.}~\bibnamefont{Schrittwieser}},
  \bibinfo{author}{\bibfnamefont{I.}~\bibnamefont{Antonoglou}},
  \bibinfo{author}{\bibfnamefont{M.}~\bibnamefont{Lai}},
  \bibinfo{author}{\bibfnamefont{A.}~\bibnamefont{Guez}},
  \bibinfo{author}{\bibfnamefont{M.}~\bibnamefont{Lanctot}},
  \bibinfo{author}{\bibfnamefont{L.}~\bibnamefont{Sifre}},
  \bibinfo{author}{\bibfnamefont{D.}~\bibnamefont{Kumaran}},
  \bibinfo{author}{\bibfnamefont{T.}~\bibnamefont{Graepel}},
  \bibnamefont{et~al.}, \bibinfo{journal}{Science}
  \textbf{\bibinfo{volume}{362}}, \bibinfo{pages}{1140} (\bibinfo{year}{2018}).

\bibitem[{\citenamefont{Mirhoseini et~al.}(2021)\citenamefont{Mirhoseini,
  Goldie, Yazgan, Jiang, Songhori, Wang, Lee, Johnson, Pathak, Nazi
  et~al.}}]{Mirhoseini1:21}
\bibinfo{author}{\bibfnamefont{A.}~\bibnamefont{Mirhoseini}},
  \bibinfo{author}{\bibfnamefont{A.}~\bibnamefont{Goldie}},
  \bibinfo{author}{\bibfnamefont{M.}~\bibnamefont{Yazgan}},
  \bibinfo{author}{\bibfnamefont{J.~W.} \bibnamefont{Jiang}},
  \bibinfo{author}{\bibfnamefont{E.}~\bibnamefont{Songhori}},
  \bibinfo{author}{\bibfnamefont{S.}~\bibnamefont{Wang}},
  \bibinfo{author}{\bibfnamefont{Y.-J.} \bibnamefont{Lee}},
  \bibinfo{author}{\bibfnamefont{E.}~\bibnamefont{Johnson}},
  \bibinfo{author}{\bibfnamefont{O.}~\bibnamefont{Pathak}},
  \bibinfo{author}{\bibfnamefont{A.}~\bibnamefont{Nazi}}, \bibnamefont{et~al.},
  \bibinfo{journal}{Nature} \textbf{\bibinfo{volume}{594}},
  \bibinfo{pages}{207} (\bibinfo{year}{2021}).

\bibitem[{\citenamefont{Shen et~al.}(2015)\citenamefont{Shen, Wang, Polson, and
  Menon}}]{Shen1:15}
\bibinfo{author}{\bibfnamefont{B.}~\bibnamefont{Shen}},
  \bibinfo{author}{\bibfnamefont{P.}~\bibnamefont{Wang}},
  \bibinfo{author}{\bibfnamefont{R.}~\bibnamefont{Polson}}, \bibnamefont{and}
  \bibinfo{author}{\bibfnamefont{R.}~\bibnamefont{Menon}},
  \bibinfo{journal}{Nature Photonics} \textbf{\bibinfo{volume}{9}},
  \bibinfo{pages}{378} (\bibinfo{year}{2015}).

\bibitem[{\citenamefont{Boutami and Fan}(2019{\natexlab{a}})}]{Boutami1:19}
\bibinfo{author}{\bibfnamefont{S.}~\bibnamefont{Boutami}} \bibnamefont{and}
  \bibinfo{author}{\bibfnamefont{S.}~\bibnamefont{Fan}}, \bibinfo{journal}{J.
  Opt. Soc. Am. B} \textbf{\bibinfo{volume}{36}}, \bibinfo{pages}{2378}
  (\bibinfo{year}{2019}{\natexlab{a}}),
  \urlprefix\url{http://www.osapublishing.org/josab/abstract.cfm?URI=josab-36-9-2378}.

\bibitem[{\citenamefont{Martin et~al.}(1994)\citenamefont{Martin, Dereux, and
  Girard}}]{Martin1:94}
\bibinfo{author}{\bibfnamefont{O.~J.~F.} \bibnamefont{Martin}},
  \bibinfo{author}{\bibfnamefont{A.}~\bibnamefont{Dereux}}, \bibnamefont{and}
  \bibinfo{author}{\bibfnamefont{C.}~\bibnamefont{Girard}},
  \bibinfo{journal}{J. Opt. Soc. Am. A} \textbf{\bibinfo{volume}{11}},
  \bibinfo{pages}{1073} (\bibinfo{year}{1994}),
  \urlprefix\url{http://www.osapublishing.org/josaa/abstract.cfm?URI=josaa-11-3-1073}.

\bibitem[{\citenamefont{Martin and Piller}(1998)}]{Martin2:98}
\bibinfo{author}{\bibfnamefont{O.~J.~F.} \bibnamefont{Martin}}
  \bibnamefont{and} \bibinfo{author}{\bibfnamefont{N.~B.}
  \bibnamefont{Piller}}, \bibinfo{journal}{Phys. Rev. E}
  \textbf{\bibinfo{volume}{58}}, \bibinfo{pages}{3909} (\bibinfo{year}{1998}),
  \urlprefix\url{https://link.aps.org/doi/10.1103/PhysRevE.58.3909}.

\bibitem[{\citenamefont{Boutami and Fan}(2019{\natexlab{b}})}]{Boutami2:19}
\bibinfo{author}{\bibfnamefont{S.}~\bibnamefont{Boutami}} \bibnamefont{and}
  \bibinfo{author}{\bibfnamefont{S.}~\bibnamefont{Fan}}, \bibinfo{journal}{J.
  Opt. Soc. Am. B} \textbf{\bibinfo{volume}{36}}, \bibinfo{pages}{2387}
  (\bibinfo{year}{2019}{\natexlab{b}}),
  \urlprefix\url{http://www.osapublishing.org/josab/abstract.cfm?URI=josab-36-9-2387}.

\bibitem[{\citenamefont{Boutami et~al.}(2020)\citenamefont{Boutami, Hassan,
  Dupré, Baud, and Fan}}]{Boutami3:20}
\bibinfo{author}{\bibfnamefont{S.}~\bibnamefont{Boutami}},
  \bibinfo{author}{\bibfnamefont{K.}~\bibnamefont{Hassan}},
  \bibinfo{author}{\bibfnamefont{C.}~\bibnamefont{Dupré}},
  \bibinfo{author}{\bibfnamefont{L.}~\bibnamefont{Baud}}, \bibnamefont{and}
  \bibinfo{author}{\bibfnamefont{S.}~\bibnamefont{Fan}},
  \bibinfo{journal}{Applied Physics Letters} \textbf{\bibinfo{volume}{117}},
  \bibinfo{pages}{071104} (\bibinfo{year}{2020}),
  \eprint{https://doi.org/10.1063/5.0013558},
  \urlprefix\url{https://doi.org/10.1063/5.0013558}.

\bibitem[{\citenamefont{Dean and Ghemawat}(2004)}]{Dean1:04}
\bibinfo{author}{\bibfnamefont{J.}~\bibnamefont{Dean}} \bibnamefont{and}
  \bibinfo{author}{\bibfnamefont{S.}~\bibnamefont{Ghemawat}}, in
  \emph{\bibinfo{booktitle}{OSDI'04: Sixth Symposium on Operating System Design
  and Implementation}} (\bibinfo{address}{San Francisco, CA},
  \bibinfo{year}{2004}), pp. \bibinfo{pages}{137--150}.

\bibitem[{\citenamefont{Akidau et~al.}(2015{\natexlab{a}})\citenamefont{Akidau,
  Bradshaw, Chambers, Chernyak, Fernández-Moctezuma, Lax, McVeety, Mills,
  Perry, Schmidt et~al.}}]{Akidau1:15}
\bibinfo{author}{\bibfnamefont{T.}~\bibnamefont{Akidau}},
  \bibinfo{author}{\bibfnamefont{R.}~\bibnamefont{Bradshaw}},
  \bibinfo{author}{\bibfnamefont{C.}~\bibnamefont{Chambers}},
  \bibinfo{author}{\bibfnamefont{S.}~\bibnamefont{Chernyak}},
  \bibinfo{author}{\bibfnamefont{R.~J.} \bibnamefont{Fernández-Moctezuma}},
  \bibinfo{author}{\bibfnamefont{R.}~\bibnamefont{Lax}},
  \bibinfo{author}{\bibfnamefont{S.}~\bibnamefont{McVeety}},
  \bibinfo{author}{\bibfnamefont{D.}~\bibnamefont{Mills}},
  \bibinfo{author}{\bibfnamefont{F.}~\bibnamefont{Perry}},
  \bibinfo{author}{\bibfnamefont{E.}~\bibnamefont{Schmidt}},
  \bibnamefont{et~al.}, \bibinfo{journal}{Proceedings of the VLDB Endowment}
  \textbf{\bibinfo{volume}{8}}, \bibinfo{pages}{1792}
  (\bibinfo{year}{2015}{\natexlab{a}}).

\bibitem[{\citenamefont{Community}(2021)}]{Apache:21}
\bibinfo{author}{\bibfnamefont{A.~B.} \bibnamefont{Community}},
  \emph{\bibinfo{title}{{Apache Beam}}} (\bibinfo{year}{2021}),
  \urlprefix\url{https://github.com/apache/beam}.

\bibitem[{\citenamefont{Akidau et~al.}(2015{\natexlab{b}})\citenamefont{Akidau,
  Bradshaw, Chambers, Chernyak, Fernández-Moctezuma, Lax, McVeety, Mills,
  Perry, Schmidt et~al.}}]{43864}
\bibinfo{author}{\bibfnamefont{T.}~\bibnamefont{Akidau}},
  \bibinfo{author}{\bibfnamefont{R.}~\bibnamefont{Bradshaw}},
  \bibinfo{author}{\bibfnamefont{C.}~\bibnamefont{Chambers}},
  \bibinfo{author}{\bibfnamefont{S.}~\bibnamefont{Chernyak}},
  \bibinfo{author}{\bibfnamefont{R.~J.} \bibnamefont{Fernández-Moctezuma}},
  \bibinfo{author}{\bibfnamefont{R.}~\bibnamefont{Lax}},
  \bibinfo{author}{\bibfnamefont{S.}~\bibnamefont{McVeety}},
  \bibinfo{author}{\bibfnamefont{D.}~\bibnamefont{Mills}},
  \bibinfo{author}{\bibfnamefont{F.}~\bibnamefont{Perry}},
  \bibinfo{author}{\bibfnamefont{E.}~\bibnamefont{Schmidt}},
  \bibnamefont{et~al.}, \bibinfo{journal}{Proceedings of the VLDB Endowment}
  \textbf{\bibinfo{volume}{8}}, \bibinfo{pages}{1792}
  (\bibinfo{year}{2015}{\natexlab{b}}).

\bibitem[{\citenamefont{Paulus et~al.}(2000)\citenamefont{Paulus, Gay-Balmaz,
  and Martin}}]{paulus:20}
\bibinfo{author}{\bibfnamefont{M.}~\bibnamefont{Paulus}},
  \bibinfo{author}{\bibfnamefont{P.}~\bibnamefont{Gay-Balmaz}},
  \bibnamefont{and} \bibinfo{author}{\bibfnamefont{O.~J.}
  \bibnamefont{Martin}}, \bibinfo{journal}{Physical Review E}
  \textbf{\bibinfo{volume}{62}}, \bibinfo{pages}{5797} (\bibinfo{year}{2000}).

\bibitem[{\citenamefont{Cheben et~al.}(2018)\citenamefont{Cheben, Halir,
  Schmid, Atwater, and Smith}}]{Cheben1:18}
\bibinfo{author}{\bibfnamefont{P.}~\bibnamefont{Cheben}},
  \bibinfo{author}{\bibfnamefont{R.}~\bibnamefont{Halir}},
  \bibinfo{author}{\bibfnamefont{J.~H.} \bibnamefont{Schmid}},
  \bibinfo{author}{\bibfnamefont{H.~A.} \bibnamefont{Atwater}},
  \bibnamefont{and} \bibinfo{author}{\bibfnamefont{D.~R.} \bibnamefont{Smith}},
  \bibinfo{journal}{Nature} \textbf{\bibinfo{volume}{560}},
  \bibinfo{pages}{565} (\bibinfo{year}{2018}).

\bibitem[{\citenamefont{Khorasaninejad
  et~al.}(2016)\citenamefont{Khorasaninejad, Chen, Devlin, Oh, Zhu, and
  Capasso}}]{Khorasaninejad:16}
\bibinfo{author}{\bibfnamefont{M.}~\bibnamefont{Khorasaninejad}},
  \bibinfo{author}{\bibfnamefont{W.~T.} \bibnamefont{Chen}},
  \bibinfo{author}{\bibfnamefont{R.~C.} \bibnamefont{Devlin}},
  \bibinfo{author}{\bibfnamefont{J.}~\bibnamefont{Oh}},
  \bibinfo{author}{\bibfnamefont{A.~Y.} \bibnamefont{Zhu}}, \bibnamefont{and}
  \bibinfo{author}{\bibfnamefont{F.}~\bibnamefont{Capasso}},
  \bibinfo{journal}{Science} \textbf{\bibinfo{volume}{352}},
  \bibinfo{pages}{1190} (\bibinfo{year}{2016}).

\bibitem[{\citenamefont{Chung and Miller}(2020)}]{Chung1:20}
\bibinfo{author}{\bibfnamefont{H.}~\bibnamefont{Chung}} \bibnamefont{and}
  \bibinfo{author}{\bibfnamefont{O.~D.} \bibnamefont{Miller}},
  \bibinfo{journal}{Optics express} \textbf{\bibinfo{volume}{28}},
  \bibinfo{pages}{6945} (\bibinfo{year}{2020}).

\bibitem[{\citenamefont{Cloud}()}]{DataflowGCP}
\bibinfo{author}{\bibfnamefont{G.}~\bibnamefont{Cloud}},
  \emph{\bibinfo{title}{{Dataflow Google Cloud}}},
  \urlprefix\url{https://cloud.google.com/dataflow}.

\bibitem[{\citenamefont{Yang et~al.}(2019)\citenamefont{Yang, Ghimire, Wu,
  Gurung, Arndt, Tsai, and Lee}}]{Yang1:19}
\bibinfo{author}{\bibfnamefont{J.}~\bibnamefont{Yang}},
  \bibinfo{author}{\bibfnamefont{I.}~\bibnamefont{Ghimire}},
  \bibinfo{author}{\bibfnamefont{P.~C.} \bibnamefont{Wu}},
  \bibinfo{author}{\bibfnamefont{S.}~\bibnamefont{Gurung}},
  \bibinfo{author}{\bibfnamefont{C.}~\bibnamefont{Arndt}},
  \bibinfo{author}{\bibfnamefont{D.~P.} \bibnamefont{Tsai}}, \bibnamefont{and}
  \bibinfo{author}{\bibfnamefont{H.~W.~H.} \bibnamefont{Lee}},
  \bibinfo{journal}{Nanophotonics} \textbf{\bibinfo{volume}{8}},
  \bibinfo{pages}{443} (\bibinfo{year}{2019}),
  \urlprefix\url{https://doi.org/10.1515/nanoph-2018-0204}.

\bibitem[{\citenamefont{Kim and Kim}(2020)}]{Kim1:20}
\bibinfo{author}{\bibfnamefont{M.}~\bibnamefont{Kim}} \bibnamefont{and}
  \bibinfo{author}{\bibfnamefont{S.}~\bibnamefont{Kim}},
  \bibinfo{journal}{Scientific Reports} \textbf{\bibinfo{volume}{10}},
  \bibinfo{pages}{1} (\bibinfo{year}{2020}).

\bibitem[{\citenamefont{Principe et~al.}(2017)\citenamefont{Principe, Consales,
  Micco, Crescitelli, Castaldi, Esposito, La~Ferrara, Cutolo, Galdi, and
  Cusano}}]{Principe1:17}
\bibinfo{author}{\bibfnamefont{M.}~\bibnamefont{Principe}},
  \bibinfo{author}{\bibfnamefont{M.}~\bibnamefont{Consales}},
  \bibinfo{author}{\bibfnamefont{A.}~\bibnamefont{Micco}},
  \bibinfo{author}{\bibfnamefont{A.}~\bibnamefont{Crescitelli}},
  \bibinfo{author}{\bibfnamefont{G.}~\bibnamefont{Castaldi}},
  \bibinfo{author}{\bibfnamefont{E.}~\bibnamefont{Esposito}},
  \bibinfo{author}{\bibfnamefont{V.}~\bibnamefont{La~Ferrara}},
  \bibinfo{author}{\bibfnamefont{A.}~\bibnamefont{Cutolo}},
  \bibinfo{author}{\bibfnamefont{V.}~\bibnamefont{Galdi}}, \bibnamefont{and}
  \bibinfo{author}{\bibfnamefont{A.}~\bibnamefont{Cusano}},
  \bibinfo{journal}{Light: Science \& Applications}
  \textbf{\bibinfo{volume}{6}}, \bibinfo{pages}{e16226} (\bibinfo{year}{2017}).

\bibitem[{\citenamefont{Asadollahbaik et~al.}(2020)\citenamefont{Asadollahbaik,
  Thiele, Weber, Kumar, Drozella, Sterl, Herkommer, Giessen, and
  Fick}}]{Asadollahbaik1:20}
\bibinfo{author}{\bibfnamefont{A.}~\bibnamefont{Asadollahbaik}},
  \bibinfo{author}{\bibfnamefont{S.}~\bibnamefont{Thiele}},
  \bibinfo{author}{\bibfnamefont{K.}~\bibnamefont{Weber}},
  \bibinfo{author}{\bibfnamefont{A.}~\bibnamefont{Kumar}},
  \bibinfo{author}{\bibfnamefont{J.}~\bibnamefont{Drozella}},
  \bibinfo{author}{\bibfnamefont{F.}~\bibnamefont{Sterl}},
  \bibinfo{author}{\bibfnamefont{A.~M.} \bibnamefont{Herkommer}},
  \bibinfo{author}{\bibfnamefont{H.}~\bibnamefont{Giessen}}, \bibnamefont{and}
  \bibinfo{author}{\bibfnamefont{J.}~\bibnamefont{Fick}}, \bibinfo{journal}{ACS
  Photonics} \textbf{\bibinfo{volume}{7}}, \bibinfo{pages}{88}
  (\bibinfo{year}{2020}),
  \eprint{https://doi.org/10.1021/acsphotonics.9b01024},
  \urlprefix\url{https://doi.org/10.1021/acsphotonics.9b01024}.

\bibitem[{\citenamefont{Phan et~al.}(2019)\citenamefont{Phan, Sell, Wang,
  Doshay, Edee, Yang, and Fan}}]{Phan1:19}
\bibinfo{author}{\bibfnamefont{T.}~\bibnamefont{Phan}},
  \bibinfo{author}{\bibfnamefont{D.}~\bibnamefont{Sell}},
  \bibinfo{author}{\bibfnamefont{E.~W.} \bibnamefont{Wang}},
  \bibinfo{author}{\bibfnamefont{S.}~\bibnamefont{Doshay}},
  \bibinfo{author}{\bibfnamefont{K.}~\bibnamefont{Edee}},
  \bibinfo{author}{\bibfnamefont{J.}~\bibnamefont{Yang}}, \bibnamefont{and}
  \bibinfo{author}{\bibfnamefont{J.~A.} \bibnamefont{Fan}},
  \bibinfo{journal}{Light: Science \& Applications}
  \textbf{\bibinfo{volume}{8}}, \bibinfo{pages}{1} (\bibinfo{year}{2019}).

\bibitem[{\citenamefont{Lin and Johnson}(2019)}]{Lin:2019}
\bibinfo{author}{\bibfnamefont{Z.}~\bibnamefont{Lin}} \bibnamefont{and}
  \bibinfo{author}{\bibfnamefont{S.~G.} \bibnamefont{Johnson}},
  \bibinfo{journal}{Optics Express} \textbf{\bibinfo{volume}{27}},
  \bibinfo{pages}{32445} (\bibinfo{year}{2019}).

\bibitem[{\citenamefont{Taflove et~al.}(2005)\citenamefont{Taflove, Hagness,
  and Piket-May}}]{Taflove1:05}
\bibinfo{author}{\bibfnamefont{A.}~\bibnamefont{Taflove}},
  \bibinfo{author}{\bibfnamefont{S.~C.} \bibnamefont{Hagness}},
  \bibnamefont{and}
  \bibinfo{author}{\bibfnamefont{M.}~\bibnamefont{Piket-May}},
  \bibinfo{journal}{The Electrical Engineering Handbook}
  \textbf{\bibinfo{volume}{3}} (\bibinfo{year}{2005}).

\bibitem[{\citenamefont{Oskooi et~al.}(2010)\citenamefont{Oskooi, Roundy,
  Ibanescu, Bermel, Joannopoulos, and Johnson}}]{Oskooi1:10}
\bibinfo{author}{\bibfnamefont{A.~F.} \bibnamefont{Oskooi}},
  \bibinfo{author}{\bibfnamefont{D.}~\bibnamefont{Roundy}},
  \bibinfo{author}{\bibfnamefont{M.}~\bibnamefont{Ibanescu}},
  \bibinfo{author}{\bibfnamefont{P.}~\bibnamefont{Bermel}},
  \bibinfo{author}{\bibfnamefont{J.~D.} \bibnamefont{Joannopoulos}},
  \bibnamefont{and} \bibinfo{author}{\bibfnamefont{S.~G.}
  \bibnamefont{Johnson}}, \bibinfo{journal}{Computer Physics Communications}
  \textbf{\bibinfo{volume}{181}}, \bibinfo{pages}{687} (\bibinfo{year}{2010}).

\end{thebibliography}
\end{document}